          [2001/04/25 1.1 (PWD)]
\documentclass[a4paper,twocolumn]{esapub} 
\usepackage{natbib,psfig}

\title{Spectral states and transient behaviour of a sample of 
 X-ray bursters observed by BeppoSax
}
\author[1]{L.Natalucci}
\author[1]{A.Bazzano}
\author[1]{F.Capitanio }
\author[1]{M.Cocchi}
\author[1]{P.Ubertini}
\author[2,3]{J.J.M. in 't Zand}
\affil[1]{IASF-CNR, via Fosso del Cavaliere 100, 00133  Roma, Italy}
\affil[2]{SRON, Sorbonnelaan 2, NL-3584
            CA Utrecht, the Netherlands
	    }
\affil[3]{
Astronomical Institute, Utrecht University, P.O.Box 80000, NL-3508 TA
            Utrecht, the Netherlands}

\begin{document}
\maketitle

\begin{abstract}
During observation campaigns of the Galactic Bulge region, BeppoSAX detected a total 
of 21 
new X-ray bursters in about 6 years of operation. These sources are mostly transient and often 
feature a hard X-ray spectrum, extending up to $\sim200$~keV. A hard spectrum is generally found 
in weak, relatively short outbursts. On the other hand two sources, SAX J1747.0-2853 and 
SAX J1750.8-2900, have been seen with soft X-ray transient behaviour. In both low/hard and 
high/soft state, a thermal component is observed at energies below ~10 keV and the spectra 
are compatible with the same model consisting of a blackbody or disk blackbody plus a hard 
or Comptonized component. Light curves are characterized by either weak, short flares or
longer, high luminosity eruptions with exponential decay.    

\end{abstract}

\section{Introduction}
High energy transient events from compact sources are characterized by rapid increases
in X-ray luminosity (greater than $\sim2$ orders of magnitude) out of a normal or quiescent state. 
Many studies in the particular topic of X-ray transients are related to accreting 
Galactic Black Holes (GBH) and to the weakly magnetized Neutron Stars (NS) in 
Low Mass X-ray Binaries (LMXB). In these sources, the X-ray outburst is originated by sudden 
increases of accretion rate probably triggered by a viscous-thermal instability in the 
accretion disk (Lasota, 2001 and refs. therein). The sources are often discovered 
in coincidence with the outbursts, as these are usually separated by very long quiescence 
periods ($\sim$~months to decades). The ASM on board RXTE and the WFC on board BeppoSAX have 
provided a wealth of new discoveries in less than a decade, resulting in a big increase 
in the number of known LMXB objects. For BeppoSAX, the systematic monitoring of the Galactic 
Bulge region as part of the mission Core Programme (see in~'t Zand et al. 2004 for review)
was complemented by a Target Of Opportunity 
(TOO) program aimed at the detailed spectral study on a wide spectral band (0.2-200 keV).  
This led to successful observations for a sample of X-ray bursters (see Table~1).         

\begin{table*}
\label{tab1} \caption{TOO of X-ray bursters performed by BeppoSAX within the Galactic 
Bulge Monitoring Program.
 }
\begin{center}
\begin{tabular}{lll}
\hline 
Source & References   &  Remarks  \\
\hline
SAX J1712.6-3739 &   in' t Zand et al. 1999a, Cocchi et al. 1999  &   \\
 &  Natalucci 2001  
   &       \\ 
SLX 1737-282 &  in' t Zand et al. 2002
   &    Detection of 15-min long burst   \\
SAX J1750.8-2900 &  Natalucci et al. 1999 + this work 
  &    Soft transient, recurrent  \\
GRS 1747-312 in Terzan 6 &  in 't Zand et al. 2000
   &     Recurrent, eclipsing source. \\ 
  &  & Detection of long, peculiar burst  \\
  &  & (in't Zand et al. 2003)
  \\ 
SAX J1810.8-2609 &   Ubertini et al.1998; Natalucci et al. 2000a
 &    Low state outburst. Hard spectrum, \\ 
 &  &  no visible cutoff.  \\
GS 1826-238 &  in' t Zand et al. 1999b
   &    Persistent source. Clocked burster \\ 
   &  & (Ubertini et al. 1999) \\ 
SAX J1747.0-2853 &  Natalucci et al. 2000b; Werner et al. 2004   
   &    Recurrent outbursts. \\ 
 & Natalucci et al. 2004 & Long activity spanning 2000-2001 \\
SAX J1748.9-2021 in NGC6440 &  In 't Zand et al. 1999c
   &       \\  
\hline
\end{tabular}
\end{center}
\end{table*}
 
\begin{figure*}[ht]
\label{fig1}
\centerline{ \vbox{
\hbox{
\psfig{figure=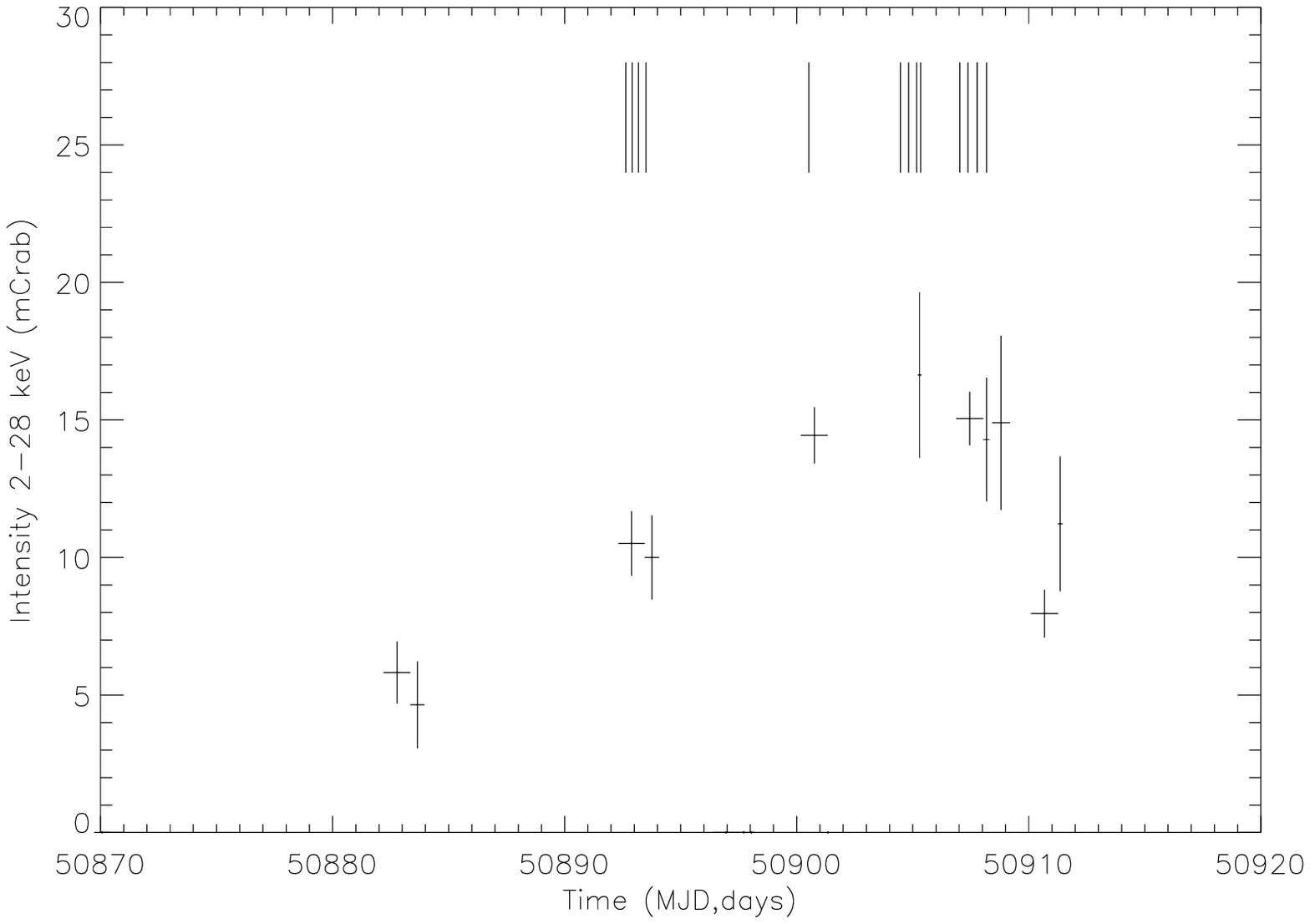,width=8cm,clip=} 
\psfig{figure=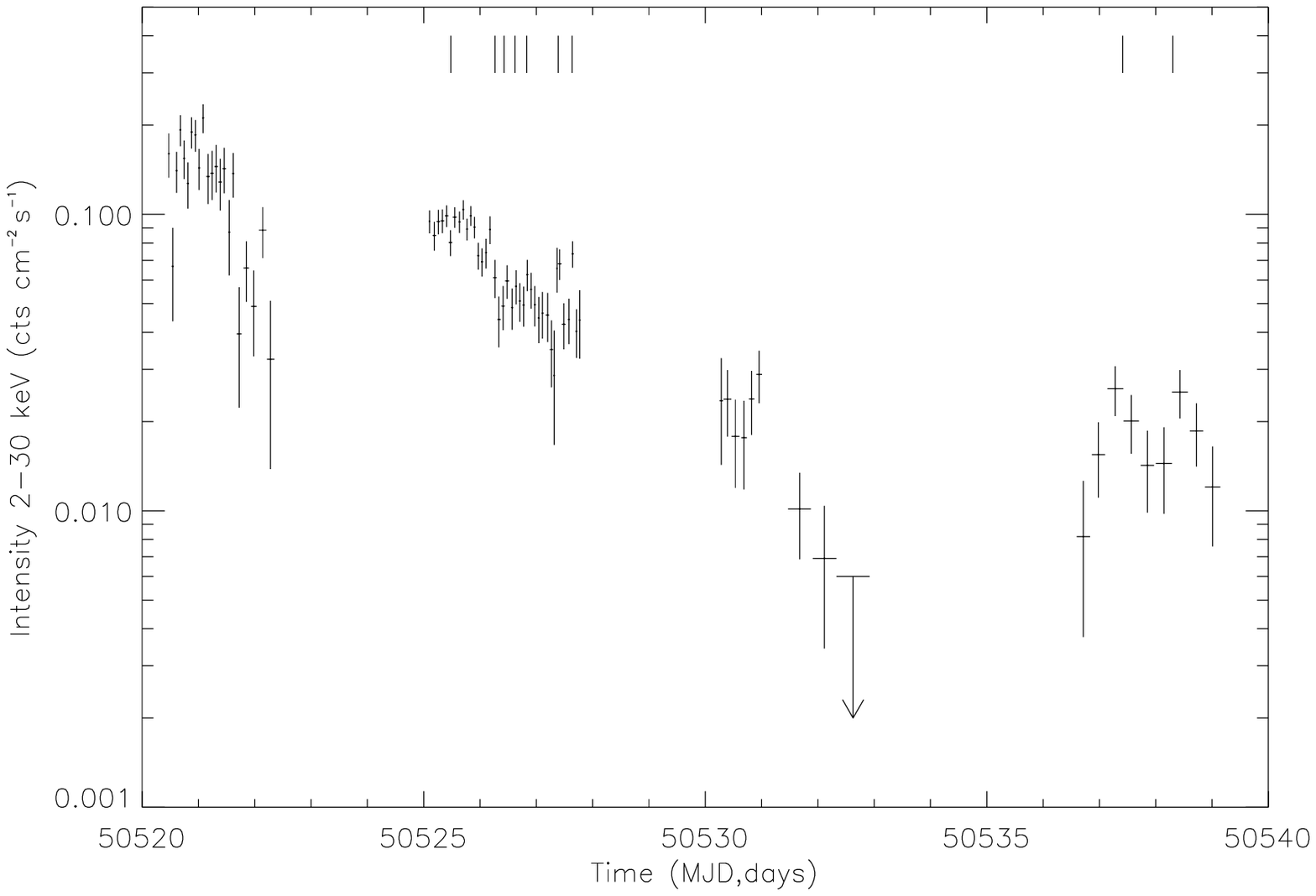,width=8cm,clip=}
}
}}

\caption{
WFC light curves of SAX J1747.0-2853 (left panel) and SAX J1750.8-2900 
(right panel) for
their discovery outbursts (March 1998 and March 1997, respectively). For both
sources, the markers indicate the occurrence time of X-ray bursts detected.
}
\end{figure*}

\section{Outburst light curves}
In transient bursters, outbursts have a typical duration of days to weeks
and type-I X-ray bursts (thermonuclear flashes on a NS surface,
lasting $\sim$~seconds to minutes) are often observed during source activity. 
As for GBH
transients, bursters may behave as soft X-ray transients (SXT) with short
rise, exponential decay light curves at high luminosities, or undergo weak
outburst (L$\leq10^{37}$~erg/s).  In Fig.1 the light curves of two transient bursters
are shown. In the case of SAX~J1747.0-2853 (1998 outburst) the flux is slowly
rising and the luminosity of the source is a small fraction ( $\leq3$\%) of the Eddington
luminosity. Both curves have been obtained with the WFC instrument
(Jager et al. 1997) on board BeppoSAX.
For SAX~J1750.8-2900 the outburst is typical of SXT and the luminosity is $10^{37}$~erg/s 
or less. For this source, an upper limit to the distance of $\sim7$~kpc
is available by the observation of the peak luminosity of X-ray bursts. 
References to these observations are given in Table~I.

Both SAX~J1747.0-2853 and SAX~J1750.8-2900 experienced further activity. 
In the spring of 2000,
SAXJ1747.8-2900 had a luminous outburst with SXT behaviour 
(Natalucci et al., 2004). A light curve
with multiple peaks was measured by RXTE/PCA (Werner et al. 2004) indicating
the presence of variations on the time scale of days. An important  
characteristic of this source is that after this long, exponential decay
outburst, the source did not return to quiescence but entered a period of
low luminosity activity (Wijnands et al 2002) for more than one year. Very 
recently (March 2004), the source brightened to an intensity greater than
200~mCrab (Markwardt \& Swank, 2004; Deluit et al., 2004).

SAX J1750.8-2900 had also a second, luminous  outburst in the spring of 2001.
During this outburst, ms oscillations in the rise of an X-ray burst and kHz QPOs
were recently discovered with RXTE (Kaaret et al. 2004). In this paper, we present
detection of further X-ray bursts and the results of a preliminary study of the
broadband spectrum obtained by the NFI (Boella et al. 1997) on board BeppoSAX.

\section{Study of broad band spectra}
Spectra of transient bursters were obtained in both low/hard and high/soft states.
Due to the low energy extension (down to $\sim0.2$~keV) of the NFI, both the thermal
emission from disk or NS and the low energy tail of the Comptonized spectrum
could be modelled.
In the low/hard state the thermal component is generally weak, contributing a
fraction of $\sim10$\% of the total flux in the 2-10 keV energy band. Modelled by
a pure blackbody, the thermal emission has a typical colour temperature below
$\sim1$~keV. The spectra of these bursters during a low/hard state outburst is
very similar to the ones of low luminosity, persistent bursters
(see e.g. Barret et al., 2000) and their high energy spectrum is generally
well described by thermal Comptonization with a plasma temperature
${kT}_{e}$~$\sim25-30$~keV. A notable exception is the source SAX~J1810.8-2900, discovered
by the WFC in March 1998, with its very hard spectrum compatible with a pure
power-law (see Fig.2, compared to the low state spectrum of SAX~J1747.0-2853). 

SAX J1747.0-2853 was also observed by the NFI in a high state. Unfortunately,
due to its vicinity to the Galactic Centre, the high energy instrument PDS was
affected by source confusion and could not be used effectively for this observation.
In Fig.3 the high state spectrum of the March 2000 outburst is shown. For this
event, the X-ray emission is characterized by a soft component with temperature
$kT\sim1.3$~keV and by a second component, which is most likely non-thermal
(see details in Natalucci et al. 2004). Since for this observation we could 
not obtain valid spectral measurements above 10.5~keV, in Fig.3 two model spectra 
are shown corresponding to spectral fits with or without a hard Comptonization 
tail.

\begin{figure}[ht]
\label{fig2}
\centerline{ \vbox{
\hbox{
\psfig{figure=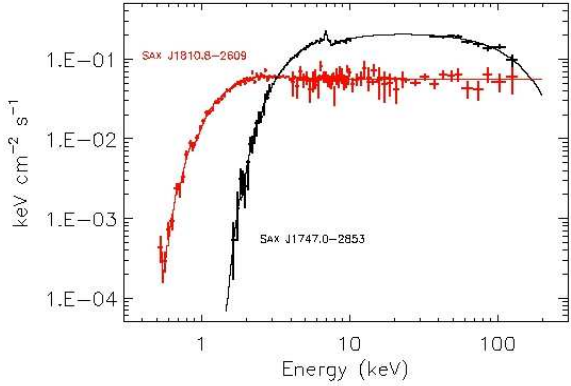,width=8cm,clip=}
}
}}

\caption{
Spectra of two X-ray bursters measured by BeppoSAX/NFI during a low
luminosity outburst. The spectrum of SAX J1810.8-2609 has no visible high
energy cutoff. References in Table~I.
}
\end{figure}

\begin{figure}[ht]
\label{fig3}
\centerline{ \vbox{
\hbox{
\psfig{figure=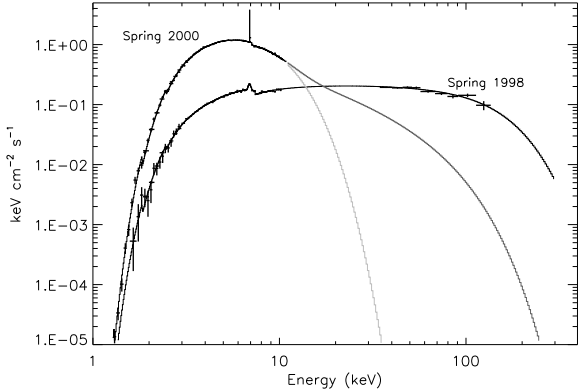,width=8cm,clip=}
}
}}

\caption{
Two different spectral states of
SAX~J1747.0-2853 (from Natalucci et al 2004). The best model spectra are shown
up to ~200 keV. See text for details.
}
\end{figure}

\section{The 2001 outburst of SAX J1750.8-2900}
After the outburst of 1997, SAX~J1750.8-2900 had a second eruption in 2001.
On this occasion, the BeppoSAX/NFI performed successfully two observations.
During the 1st one on 23-24 March, the source intensity was $\sim4$~mCrab (2-10 keV)
and the source was in a rising phase; on 9-10 April, the intensity was slowly
decaying with an average value of 43 mCrab. The light curve of this observation
in shown in Fig.4 for the MECS instrument (units 2+3). In this dataset we detect
the presence of three X-ray bursts, which are visible in the 100~s binned curve.

\begin{figure}[ht]
\label{fig4}
\centerline{ \vbox{
\hbox{
\psfig{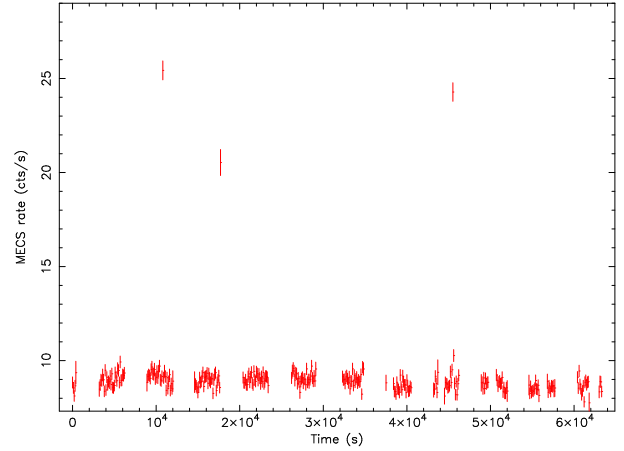}
}
}}

\caption{
MECS light curve of SAX~J1750.8-2900, binned on 100s. The three
spikes are due to the occurrence of X-ray bursts.
}
\end{figure}

\begin{figure}[ht]
\label{fig5}
\centerline{ \vbox{
\hbox{
\psfig{figure=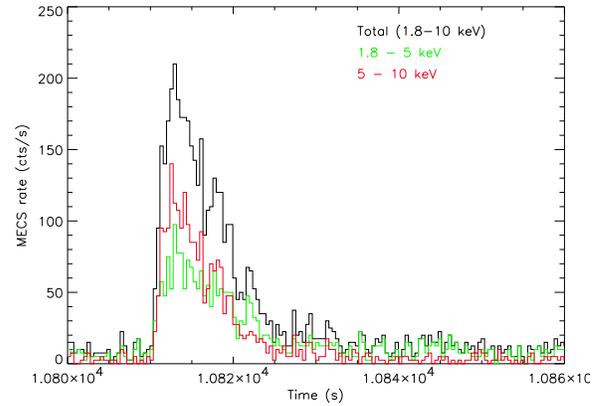,angle=90,width=8cm,clip=}
}
}}

\caption{
Time profile of the first detected burst, in two different energy
bands. The spectral softening typical of these thermonuclear flash events is evident.
}
\end{figure}

\begin{figure*}[ht]
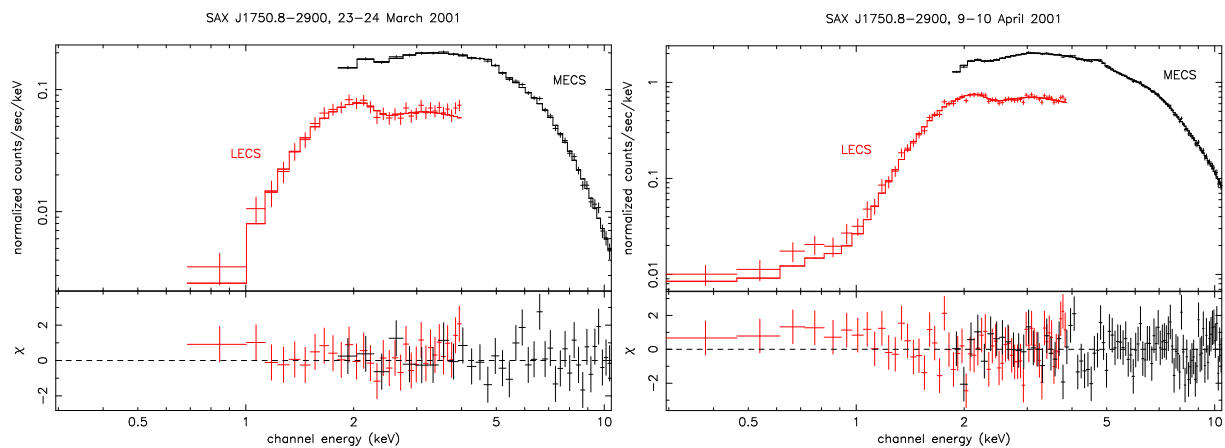

\label{fig6}
\centerline{ \vbox{
\hbox{
\psfig{figure=monaco_fig6a.ps,angle=-90,width=8cm,clip=}
\psfig{figure=monaco_fig6b.ps,angle=-90,width=8cm,clip=}
}
}}

\caption{
NFI count rate spectra for the two observations of SAX J1750.8-2900, performed
during the most recent outburst of March-April 2001.
The two observations are separated by ~2 weeks, showing a factor of $\sim10$ difference
in intensity.
}
\end{figure*}

In Fig.5 the profile of the first X-ray burst is shown for two different energy
bands. From the plot itself we see clear evidence of spectral softening 
with time (as expected from a
type-I X-ray burst). The intensity of this burst corresponds to a bolometric
fluence of $\sim10^{-7}$~erg~cm$^{-2}$~s$^{-1}$, similar to the other 
bursts observed from this source.
No X-ray bursts were observed during the first observation, when the intensity
was 10 times lower. This is reminiscent of the behaviour observed in 1997, where
most X-ray bursts occurred at an intermediate level of persistent flux 
(see Fig. 1, right panel).

In Fig. 6 the count rate spectrum is shown for the two NFIs, after subtraction
of the data sections corresponding to the detected bursts. Both observations
were fitted with a model consisting of two components: a multicolor disk
blackbody plus thermal Comptonization. A standard systematic error of 
1\% was used to account for calibration related uncertainties. The above model is 
found to provide good fits: 
$\chi^{2}_{r}$=0.84 (61 dof)
and $\chi^{2}_{r}$=1.07 (144 dof) for the 1st and 2nd observation, respectively. For
the second observation, the model is in good agreement with the spectrum found
by RXTE/PCA  (Kaaret et al. 2002) in a quasi-simultaneous observation, yielding
a relatively low plasma temperature (${kT}_{e}$~$\sim5.5$~keV in our fit). However, the temperature
is not well constrained. We attempted to add the PDS data (15-200 keV) but due
to possible significant source contamination, this analysis is difficult and 
work is still in progress.

\section{Conclusions}
A significant sample of the X-ray bursters detected by the BeppoSAX WFC during the Galactic Centre 
Monitoring program has been studied by dedicated follow-up observations 
with the NFI instruments. One of the
sources (SAX~J1747.0-2853) had outbursts in both low/hard and high/soft state besides 
showing very long ($\sim1$~year or more), low luminosity activity. Another recurrent
transient, SAX~J1750.8-2900, showed two relatively luminous episodes with SXT 
behaviour, the latest occurring in March 2001. In this paper we have presented the main spectral 
characteristics of these sources along with some preliminary results of 
the second outburst of SAX~J1750.8-2900. During this event, at least three X-ray bursts 
were present in the light curve, 
which were clearly spot by the sensitive NFIs.

\section*{acknowledgements}

The BeppoSAX satellite is a joint Italian and
Dutch programme. We thank ASI for the prolonged, continuous handling of the
mission and operations. 
LN is grateful to M.~Federici for providing valuable  
technical support.

\end{document}